# Efficient Approximation for Triangulation of Minimum Treewidth

Eyal Amir
Department of Computer Science,
Gates Building, 2A wing
Stanford University, Stanford, CA 94305-9020, USA
eyal.amir@cs.stanford.edu

## Abstract

We present four novel approximation algorithms for finding triangulation of minimum treewidth. Two of the algorithms improve on the running times of algorithms by Robertson and Seymour, and Becker and Geiger that approximate the optimum by factors of 4 and $3\frac{2}{3}$, respectively. A third algorithm is faster than those but gives an approximation factor of $4\frac{1}{2}$. The last algorithm is yet faster, producing factor-$O(lgk)$ approximations in polynomial time. Finding triangulations of minimum treewidth for graphs is central to many problems in computer science. Real-world problems in artificial intelligence, VLSI design and databases are efficiently solvable if we have an efficient approximation algorithm for them. We report on experimental results confirming the effectiveness of our algorithms for large graphs associated with real-world problems.

## 1 INTRODUCTION

Given an undirected graph, $G$, and an integer $k$, TREEWIDTH is the problem of deciding whether the treewidth of $G$ is at most $k$ [Robertson and Seymour, 1986]. An equivalent constructive problem is finding a *triangulation* of $G$ with a clique number that is at most $k + 1$ (the clique number of a graph is the size of the largest clique in this graph). This is also equivalent to finding a *tree decomposition* or a *junction tree* of $G$ with width at most $k$.

An efficient solution to this problem is key in many applications in artificial intelligence, databases and logical-circuit design. Exact inference in Bayesian networks using the junction tree algorithm [Lauritzen and Spiegelhalter, 1988, Jensen et al., 1990] requires us to first find a junction tree and then perform inference using that tree. The time complexity of the junction tree algorithm depends exponentially on the width of the tree, so it is important to try to find a close to optimal clique tree. Reasoning with structured CSPs, propositional SAT and FOL problems also benefits from efficiently finding close-to-optimal tree decompositions [Dechter and Pearl, 1989, Amir and McIlraith, 2000]. The solution time of many graph-related NP-hard problems is possible in polynomial time if the graph has low treewidth and a triangulation of minimum treewidth is given (e.g., [Arnborg et al., 1991]).

In this paper we present four approximation algorithms for finding triangulations of minimum treewidth. The first algorithm improves an algorithm of [Robertson and Seymour, 1995] and produces factor-4 approximations in time $O(2^{4.38k}n^2k)$, where $n, k$ are the number of nodes and the treewidth of the given graph, $G$, respectively. The second algorithm improves an algorithm of [Becker and Geiger, 1996] and produces factor-$3\frac{2}{3}$ approximations in time $O(2^{3.6982k}n^3k^3lg^4n)$. The third algorithm produces factor-$4\frac{1}{2}$ triangulations in time $O(2^{3k}n^2k^{\frac{3}{2}})$. The last algorithm produces a factor-$O(lgk)$ approximation in time $O(n^3lg^4nk^5\,lgk)$.

The time bounds achieved by the first and second algorithms are faster by factors of $O(2^{0.4k})$ and $O(2^kpoly(n))$, respectively, than previously available algorithms for these approximation factors. The third algorithm has faster combined $n, k$ time than any known algorithm that produces a constant-factor approximation. The last algorithm is the first polynomial-time algorithm that approximates the optimal by a factor that does not depend on $n$. We have implemented the 4-approximation and $4\frac{1}{2}$-approximation algorithms and used them to find tree decompositions of graphs used in the a subset of the HPKB project [Cohen et al., 1999], a subset of the CYC knowledge base [Lenat, 1995], and several CPCS Bayesian networks [Pradhan et al., 1994]. These graphs have between 100 and 600 nodes and between 400 and 4000 edges. Our results compare favorably with the algorithms of [Becker and Geiger, 1996, Shoikhet and Geiger, 1997].

Optimally solving TREEWIDTH is known to be NP-hard [Arnborg et al., 1987], and so is the closely related op-timal decomposition of Bayesian Networks [Wen, 1990]. It is an open question whether a constant-factor approximation can be found in polynomial time. Nevertheless, several algorithms with guaranteed optimal solutions (e.g., [Bodlaender, 1996, Shoikhet and Geiger, 1997]) or constant-factor approximations to the optimal (e.g., [Robertson and Seymour, 1995, Reed, 1992, Kloks, 1994, Becker and Geiger, 1996]) were found. These algorithms take time that depends polynomially on $n$ but exponentially on $k$, the treewidth of the graph. Most of them cannot solve TREEWIDTH for graphs of treewidth greater than 4 in any reasonable amount of time. (e.g., see [Röhrig, 1998]). The best approximation guarantee in polynomial time is due to [Kloks, 1994, Bodlaender et al., 1995] who achieved a $O(log(n) \cdot k)$-factor approximation.

Algorithms with the best time bounds found so far that do not assume a bounded treewidth are due to [Becker and Geiger, 1996] (factor-$3\frac{2}{3}$ approximation, with time $O(2^{4.66k} \cdot n \cdot poly(n))$, where $poly(n)$ is the running time of linear programming), and [Reed, 1992] (factor-5 approximation, with time $O(3^{4k}k^2 n \, lgn))$. Experiments with the fastest algorithms available show that TREEWIDTH of graphs with treewidth of 10 or more cannot be solved in any reasonable amount of time (reasonable here is less than 24 hours), for graphs of 100 nodes or more (see [Becker and Geiger, 1996, Shoikhet and Geiger, 1997]). Graphs of sizes larger than these are exactly those that are of interest in many of the above-mentioned applications.

Section 2 defines the main notions involved in computing treewidth and recalls some theorems proved elsewhere. Section 3 presents our 4- and $4\frac{1}{2}$- approximation algorithms. Section 4 presents our $3\frac{2}{3}$- and $O(lgk)$- approximation algorithms. The paper concludes with experimental results.

The algorithms in this paper are described for constant-weight nodes (applicable to binary nodes in Bayesian Networks). Extensions for weighted nodes are possible along similar lines. A good survey paper on TREEWIDTH is [Bodlaender, 1997].

## 2 TREEWIDTH

In this section we briefly recall some of the main definitions pertaining to treewidth.

A cycle in a graph is *chordless* if no proper subset of the vertices of the cycle forms a cycle.

**Definition 2.1** *A graph is* triangulated *(or* chordal*) if it contains no chordless cycle of length greater than three.*

A *triangulation* of a graph $G$ is a graph $H$ with the same set of vertices such that $G$ is a subgraph of $G$ and such that $H$ is triangulated.

**Definition 2.2 ([Robertson and Seymour, 1986])** *A tree-decomposition of a graph $G(V, E)$ is a pair $D = (S, T)$ with $S = \{X_i \mid i \in I\}$ a collection of subsets of vertices of $G$ and $T = (I, F)$ a tree, with one node for each subset of $S$, such that the following three conditions are satisfied:* (1) $\bigcup_{i \in I} X_i = V$. (2) *For all edges* $(v, w) \in E$ *there is a subset $X_i \in S$ such that both $v, w$ are contained in $X_i$.* (3) *For each vertex $x$, the set of nodes $\{i \mid x \in X_i\}$ forms a subtree of $T$.*

The *width* of a *tree-decomposition* $(\{X_i \mid i \in I\}, T = (I, F))$ is $max_{i \in I}(|X_i| - 1)$. The *treewidth* of a graph $G$ equals the minimum width over all tree-decompositions of $G$. Equivalently, the treewidth of $G$ is the minimum $k \geq 0$ such that $G$ is a subgraph of a triangulated graph with all cliques of size at most $k + 1$. Any triangulation of a graph defines a tree-decomposition of a graph of the same treewidth. Similarly, every tree-decomposition of a graph defines a triangulation of it of the same treewidth.

**Definition 2.3** *Let $G(V, E)$ be a graph, $W \subseteq V$ a subset of the vertices and $\alpha \in (0, 1)$ a real number. An $\alpha$-vertex-separator of $W$ in $G$ is a set of vertices $X \subseteq V$ such that every connected component of $G[V \setminus X]$ has at most $\alpha|W|$ vertices of $W$. A two-way $\alpha$-vertex-separator is required in addition to have exactly two sets, $S_1, S_2$, separated by $X$ such that $S_1 \cup S_2 \cup X = V$ and $|S_i| \leq \alpha|W|, i = 1, 2$.*

**Lemma 2.4 ([Robertson and Seymour, 1986])** *Let $G(V, E)$ be a graph with $n$ vertices and treewidth $k$. There exists a set $X$ with $k + 1$ vertices such that every connected component of $G[V \setminus X]$ has at most $\frac{1}{2}(n - k)$ vertices.*

**Corollary 2.5 ([Becker and Geiger, 1996])** *Let $G(V, E)$ be a graph with $n \geq k + 1$ vertices and treewidth $k$. For every $W \subseteq V, |W| > 1$, there is a vertex separator $X$ and sets $A, B, C \subset V$ such that $A \cup B \cup C \cup X = V$, $A, B, C$ are separated by $X$, $|X| \leq k$ and $|W \cap C| \leq |W \cap B| \leq |W \cap A| \leq \frac{1}{2}|W|$.*

## 3 USING 2-WAY VERTEX SEPARATORS

The two algorithms presented in this section use two-way separators recursively. They differ on their choice of actual separator: 2/3 versus 1/2.

### 3.1 MINIMUM VERTEX SEPARATORS

We briefly describe the notion of a vertex separator. Let $G = (V, E)$ be an undirected graph. A set $S$ of vertices is called an $(a, b)$-*vertex-separator* if $\{a, b\} \subset V \setminus S$ and every path connecting $a$ and $b$ in $G$ passes through at least one vertex contained in $S$. An $(a, b)$-vertex-separator of minimum cardinality is said to be a *minimum $(a, b)$-vertex-separator*. The weaker property of a vertex separator be-ing *minimal* requires that no subset of the $(a, b)$-vertex-separator is an $(a, b)$-vertex-separator.

Algorithms for finding minimum vertex separators typically reduce the problem to a maximum flow problem in a directed graph. The algorithm of Even and Tarjan reported in [Even, 1979] for finding minimum vertex separators uses Dinitz's algorithm [Dinic, 1970] with time complexity $O(|V|^{\frac{1}{2}}|E|)$.

Another possibility is to use the Ford-Fulkerson flow algorithm [Ford Jr. and Fulkerson, 1962] (alternatively, see [Cormen et al., 1989]), for computing maximum flow. For an original graph of treewidth $< k$ this involves finding at most $k$ augmenting paths of capacity 1. Thus, the combined algorithm using the Ford-Fulkerson maximum flow algorithm finds a minimum $(a, b)$-vertex-separator in time $O(k(|V| + |E|))$.

## 3.2 FACTOR-4 APPROXIMATION ALGORITHM

Procedure *2way-2/3-triang*, displayed in Figure 1, finds factor-4 approximations. For a graph $G$ and a parameter $k$, running 2way-2/3-triang($G, \emptyset, k$), either returns a valid answer that the the treewidth of $G$ is of size $> k - 1$ or it returns a triangulation of $G$ of clique number at most $4k+1$.

---

PROCEDURE 2way-2/3-triang($G, W, k$)
$G = (V, E)$ with $|V| = n$, $W \subseteq V$, $k$ integer.

1. If $n \leq 4k$, then make a clique of $G$. Return.

2. Let $W' \leftarrow W$. Add to $W'$ vertices from $V$ such that $|W'| = 3k + 2$.

3. Find $X$, a minimum $\frac{2}{3}$-vertex-separator of $W'$ in $G$, with $S_1, S_2$ two nonempty parts separated by $X$ ($S_1 \cup S_2 \cup X = V$) and $|X| \leq k$. If there is no such separator, then output "*the treewidth exceeds $k - 1$*" and exit.

4. For $i \leftarrow 1$ to 2 do
   (a) $W_i \leftarrow S_i \cap W$.
   (b) call 2way-2/3-triang($G[S_i \cup X], W_i \cup X,$ k).

5. Add edges between vertices of $W \cup X$, making a clique of $G[W \cup X]$.

---

Figure 1: A factor-4 approximate triangulation algorithm.

This algorithm is very similar to that of [Robertson and Seymour, 1995], as presented in [Reed, 1992]. The main difference is the more efficient algorithm that we use for exact vertex separation, which we provide below. The addition of elements to $W'$ in step 2 ensures completeness of our separator (see Lemma 3.2's proof).

**Lemma 3.1** *If $G(V, E)$ is a graph, $k$ an integer and $W \subseteq V$ such that $|W| \leq 3k + 2$, then 2way-2/3-triang($G,W,k$) either outputs correctly that the treewidth of $G$ is more than $k$ or it triangulates $G$ such that the vertices of $W$ form a clique and the clique number of the result is at most $4k + 1$.*

The proof is identical to that presented in [Robertson and Seymour, 1986, Reed, 1992].

Figure 2 presents the algorithm we will use for finding a $\frac{2}{3}$-vertex-separator of $W'$ in $G$ (step 3 in procedure *2way-2/3-triang*). It checks choices of sets of vertices to be separated until a solution is found or the choices are exhausted. The intuition behind making a clique from each selected set, $W^i$, is that doing so prevents any element from that clique from becoming an element in the separated subset of the other side. Given an arbitrary vertex separator of $v_{W^1}, v_{W^2}$, any vertex in the clique of $W^1$ must be either in the separator itself or in $S_1$.

---

PROCEDURE $\frac{2}{3}$-vtx-sep($W, G, k$)
$G = (V, E)$ with $|V| = n$, $W \subseteq V$, $k$ integer.

1. Nondeterministically take a set $W^1$ of $\lceil \frac{|W|}{2} \rceil$ vertices from $W$ and a set $W^2$ of $\lceil \frac{|W|}{3} \rceil$ vertices from $W \setminus W^1$.

2. Let $G' \leftarrow G$. Add edges to $G'$ so that $W^1$ is a clique and $W^2$ is a clique. Create new vertices $v_{W^1}, v_{W^2}$ in $G'$ and connect them to all the vertices of $W^1, W^2$, respectively.

3. Find a minimum $(v_{W^1}, v_{W^2})$-vertex-separator, $X$. If $|X| \leq k$, return $|X|$ and two separated subsets $S_1, S_2$, discarding $v_{W^1}, v_{W^2}$. Otherwise, return "*failure*".

---

Figure 2: Find a $\frac{2}{3}$-vertex-separator of $W$ in $G$.

**Lemma 3.2** *Let $G(V, E)$ be a graph, $k \geq 0$ an integer, and $W \subseteq V$ of size $3k + 2$. Algorithm $\frac{2}{3}$-vtx-sep($W$, $G$, $k$) finds a $\frac{2}{3}$-separator of $W$ in $G$ of size $\leq k$, if it exists, returning* failure *otherwise. It does so in time $O(\frac{2^{4.38k}}{k} f(|V|, |E| + k^2, k))$, given a min-$(a, b)$-vertex-separator algorithm taking time $f(n, m, k)$.*

PROOF    We prove the correctness of the algorithm first. Assume that the algorithm finds a separator $X$ of $S_1, S_2$ in $G'$. $X$ is also a separator of $S_1, S_2$ in $G$, by the way we constructed $G'$ from $G$. Also, $X$ separates $W^1 \setminus X$ and $W^2 \setminus X$ in $G'$ because $W^1 \cup \{v_{W^1}\}$ and $W^2 \cup \{v_{W^2}\}$ are cliques in $G'$ and $X$ separates $v_{W^1}, v_{W^2}$ (if $X$ does not separate $W^1 \setminus X$ and $W^2 \setminus X$ in $G'$, then there is a path between $v_{W^1}, v_{W^2}$ that does not go through $X$). Finally, $X$ is a $\frac{2}{3}$-vertex-separator of $W$ because $|W^1|, |W^2| \geq \frac{|W|}{3}$, $W^1 \setminus X \subset S_1$ and $W^2 \setminus X \subseteq S_2$, so $|S_i \cap W| \leq \frac{2}{3}|W|$, for $i = 1, 2$. Notice that $S_1, S_2$ are never empty because $|X| \leq k$ and $|S_i| \geq |W^i| - |X| \geq 1$ for $i = 1, 2$ ($|W^i| \geq \lceil \frac{|W|}{3} \rceil = k + 1$ because $|W| = 3k + 2$).

For the reverse direction, assume that the treewidth of $G$ is $k - 1$ and we show that the algorithm will find a suitable separator. Assume first that there are two sets of vertices $S_1, S_2$ separated by $X$ in $G$ such that $S_1 \cup S_2 \cup X = V$ and $|S_i \cap W| \le \frac{|W|}{2}$, for $i = 1, 2$. Let $W_{\text{n\_sep}}^i = W \cap S_i$, for $i = 1, 2$. Let $W_{\text{sep}}^i \subseteq W \cap X$ such that $W_{\text{sep}}^1 \cup W_{\text{sep}}^2 = W \cap X$ and $|W_{\text{sep}}^i \cup W_{\text{n\_sep}}^i| = \frac{|W|}{2}$. Let $W^i = W_{\text{sep}}^i \cup W_{\text{n\_sep}}^i$, for $i = 1, 2$. Then, $X$ separates $W^1 \setminus X, W^2 \setminus X$, as $W^i \setminus X = W_{\text{n\_sep}}^i$, for $i = 1, 2$. Thus, running steps 2,3 in our algorithm using this selection of $W^1, W^2$ will find a separator of size $\le |X| \le k$. By the previous paragraph, this separator is a $\frac{2}{3}$-vertex-separator of $W$ in $G$.

Now we show that if there are no such sets $S_1, S_2, X$, then our algorithm still finds a suitable separator. By Lemma 2.5, there are three sets, $A, B, C$, of vertices separated by $X$ in $G$ such that $|X| \le k$ and $|W \cap C| \le |W \cap B| \le |W \cap A| \le \frac{1}{2}|W|$. Let $S_1 = A, S_2 = B \cup C$. If $|S_2 \cap W| \le \frac{|W|}{2}$, then the first selection case would cover this $W$ (the previous paragraph). Thus, $|S_2 \cap W| > \frac{|W|}{2}$. Take $W^1 \subset (S_2 \cap W)$ of size $\frac{|W|}{2}$ and $W^2 \subset ((S_1 \cup X) \cap W) \setminus W^1$ of size $\frac{|W|}{3}$. The selection of $W^2$ is possible because $|(S_1 \cup X) \cap W| = |S_1 \cap W| + |X \cap W| \ge \frac{1}{3}|W \setminus X| + |X \cap W| = \frac{1}{3}|W|$. For this selection of $W^1, W^2$ our algorithm will find a separator of size $\le |X| \le k$ because $X$ is already a separator of $W^1, W^2 \setminus X$. By the first paragraph in this proof, this separator is a $\frac{2}{3}$-vertex-separator of $W$ in $G$.

Finally, each choice of $W^1$ takes $O(f(|V|, |E| + k^2, k))$ time to check, for $f(n, m, k)$ the time taken by a $min$-$(a,b)$-$vertex$-$separator$ algorithm over a graph with $n$ vertices, $m$ edges and treewidth $k-1$. There are $\binom{3k+2}{1.5k+1}$ ways to choose $1.5k+1$ elements ($W^1$) from a set of $3k+2$ elements ($W$). Also, there are $\binom{1.5k+1}{k+1}$ ways to choose $k+1$ elements ($W^2$) from a set of $1.5k+1$ elements ($W \setminus W^1$). Since $\binom{3k+2}{1.5k+1} = O(\frac{2^{3k}}{\sqrt{k}})$ and $\binom{1.5k+1}{k+1} = O(\frac{2^{1.3776k}}{\sqrt{k}})$ (using Stirling's approximation), we get the time bound of $O(\frac{2^{4.3776k}}{k} f(|V|, |E| + k^2, k))$. ∎

**Proposition 3.3 (cf [Reed, 1992])** *If the treewidth of $G(V,E)$ is $k-1$, then $|E| \le |V|k$.*

**Theorem 3.4** *Procedure 2way-2/3-triang(G, ∅, k) finds a triangulation of $G$ of clique number $\le 4k+1$, if the treewidth of $G$ is at most $k-1$, in time $O(2^{4.38k}|V|^{\frac{5}{2}})$ or $O(2^{4.38k}|V|^2 k)$ if we use the minimum $(a,b)$-vertex-separator algorithm of [Even, 1979] or [Ford Jr. and Fulkerson, 1962], respectively.*

PROOF Lemmas 3.1 and 3.2 prove the correctness. For the time bound, [Reed, 1992] showed that there are $O(|V|)$ recursive calls to such triangulation algorithms. Since each recursive step runs $\frac{2}{3}$-$vtx$-$sep$ once and makes a clique of size $\le 4k+2$, we get that the combined procedure using [Even, 1979]'s algorithm for min-$(a,b)$-vertex-separator (time $O(|V|^{\frac{1}{2}}|E|)$) takes time $O(\frac{2^{4.38k}}{k}|V|^{\frac{1}{2}}(|E|+k^2)|V|)$. Using Proposition 3.3 we get the bound $O(\frac{2^{4.38k}}{k}|V|^{\frac{3}{2}}(|V|k+k^2)) = O(2^{4.38k}|V|^{\frac{5}{2}})$. Similarly, using the algorithm given by [Ford Jr. and Fulkerson, 1962] for finding a minimum $(a,b)$-vertex-separator in time $O(k(n+m))$ we get time $O(2^{4.38k}|V|^2 k)$. ∎

### 3.3 FACTOR-$4\frac{1}{2}$ APPROXIMATION ALGORITHM

We can avoid many of the choices examined in procedure $\frac{2}{3}$-vtx-sep if we allow the resulting separator to be slightly larger. Procedure 2way-half-vtx-sep, presented in Figure 4 does that, returning a minimum two-way $\frac{1}{2}$-vertex separator. The combined procedure, called *2way-half-triang*, is identical to procedure 2way-2/3-triang besides replacing step 3. It is presented in Figure 3.

---

PROCEDURE 2way-half-triang($G, W, k$)
$G = (V, E)$ with $|V| = n$, $W \subseteq V$, $k$ integer.

1. If $n \le 4k$, then make a clique of $G$. Return.

2. Let $W' \leftarrow W$. Add to $W'$ vertices from $V$ such that $|W'| = 3k + 2$.

3. Find $X$, a minimum two-way $\frac{1}{2}$-vertex-separator of $W'$ in $G$, with $S_1, S_2$ the two nonempty parts separated by $X$ ($S_1 \cup S_2 \cup X = V$) and $|X| \le 1\frac{1}{2}k$. If there is no such separator, then output "*the treewidth exceeds $k-1$*" and exit.

4. For $i \leftarrow 1$ to 2 do
   (a) $W_i \leftarrow S_i \cap W$.
   (b) call 2way-half-triang($G[S_i \cup X], W_i \cup X$, k).

5. Add edges between vertices of $W \cup X$, making a clique of $G[W \cup X]$.

---

Figure 3: A factor-$4\frac{1}{2}$ approximate triangulation algorithm.

**Lemma 3.5** *If $G(V,E)$ is a graph with treewidth $< k$ and $W \subseteq V$, then there is a two-way $\frac{1}{2}$-vertex-separator of $W$ in $G$ with size at most $k + \frac{1}{6}|W|$*

PROOF By Lemma 2.5 there are $A, B, C \subset V$ separated by $X$ such that $A \cup B \cup C \cup X = V$, $|X| \le k$ and $|W \cap C| \le |W \cap B| \le |W \cap A| \le \frac{1}{2}|W|$. If $|(B \cup C) \cap W| \le \frac{1}{2}|W|$, then $A, (B \cup C)$ and $X$ satisfy our desired conditions.

Thus, assume that $|(B \cup C) \cap W| > \frac{1}{2}|W|$. Take $X_C \subset W \cap C$ of size $|(B \cup C) \cap W| - \frac{1}{2}|W|$. Then $|X_C| = |(B \cup C) \cap W| - \frac{1}{2}|W| \le \frac{2}{3}|W| - \frac{1}{2}|W| = \frac{1}{6}|W|$. Let $X' = X \cup X_C$, $S_1 = A$ and $S_2 = (B \cup C) \setminus X_C$. This $X', S_1, S_2$ satisfy the desired conditions because $|S_2 \cap W| \le \frac{1}{2}|W|$, $|S_1| \le \frac{1}{2}|W|$, $|X'| \le |X| + |X_C| \le k + \frac{1}{6}|W|$ and $X'$ separates $S_1, S_2$ (because $X$ separates $S_1, S_2$). ∎

**Lemma 3.6** *If $G(V,E)$ is a graph with $n$ vertices, $k$ an integer and $W \subseteq V$ such that $|W| \le 3k + 2$, then 2way-half-triang($G, W, k$) either outputs correctly that the treewidth* of $G$ is more than $k-1$ or it triangulates $G$ such that the vertices of $W$ form a clique and the clique number of the resulting graph is at most $4\frac{1}{2}k + 2$.

PROOF If the algorithm outputs that the treewidth is more than $k-1$, then it did not find a decomposition of $W$ as needed. If the treewidth is at most $k-1$, then Lemma 3.5 guarantees the existence of a two-way $\frac{1}{2}$-vertex-separator of $W$ in $G$ with size at most $k + \frac{1}{6}|W|$. Thus, this separator is of size at most $k + \frac{1}{6}|W| \leq k + \frac{1}{6}(3k+2) = 1\frac{1}{2}k + \frac{1}{3}$ (and because the size cannot be fractional, it is at most $1\frac{1}{2}k$). If we did not find such a separator, then the treewidth is indeed at most $k-1$.

The same argument used for the proof of Lemma 3.1 shows that the algorithm always terminates and, if it is successful, then it returns a graph that is triangulated.

We show that the clique number of this triangulation is at most $4\frac{1}{2}k + 2$. First, notice that always $|W| \leq 3k+2$. Initially, $|W| \leq 3k+2$ by our assumption in the statement of the lemma. As the algorithm is called recursively, $|X| \leq 1\frac{1}{2}k$ and $|W_i| \leq \frac{1}{2}|W'| = 1\frac{1}{2}k + 1$. Thus, $|W_i \cup X| \leq 1\frac{1}{2}k + 1 + 1\frac{1}{2}k = 3k+1$, which concludes the induction step ($W$ in the recursive call to the algorithm is $W_i \cup X$).

Now, let $M$ be a maximal clique. If $M$ contains no vertex of $S_i \setminus W_i$, for $i = 1, 2$, then $M$ contains only vertices of $W \cup X$. Thus, $|M| \leq 3k + 2 + 1\frac{1}{2}k = 4\frac{1}{2}k + 2$. On the other hand, if $M$ contains a vertex of $S_i \setminus W_i$, then it does not contain any vertex of $S_j$, for $j \neq i$. This is because $X$ vertex-separates $S_1, S_2$ (any two separated vertices cannot have an edge connecting them). Hence, $M$ is a clique in the triangulation of $G[S_i \cup X]$. By induction we know that $|M| \leq 4\frac{1}{2}k + 2$. This proves the lemma. ∎

Procedure 2way-half-vtx-sep is very similar to procedure $\frac{2}{3}$-vtx-sep with one main difference. While $\frac{2}{3}$-vtx-sep selects two sets of sizes $\frac{1}{2}|W|$ and $\frac{1}{3}|W|$, procedure 2way-half-vtx-sep selects two sets of size $\frac{1}{2}|W|$. This precludes finding two-way separators in which one of the sets is of size $\frac{2}{3}|W|$ (as we did before).

---

PROCEDURE 2way-half-vtx-sep($W, G, k$)
$G = (V, E)$ with $|V| = n$, $W \subseteq V$, $k$ integer.

1. Nondeterministically choose a set $W^1$ of $\frac{|W|}{2}$ vertices from $W$. Let $W^2$ be $W \setminus W^1$.

2. Let $G' \leftarrow G$. Add edges to $G'$ so that $W^1$ is a clique and $W^2$ is a clique. Create new vertices $v_{W^1}, v_{W^2}$ in $G'$ and connect them to all the vertices of $W^1, W^2$, respectively.

3. Find a minimum $(v_{W^1}, v_{W^2})$-vertex-separator, $X$. If $|X| \leq 1\frac{1}{2}k$, return $|X|$ and two separated subsets $S_1, S_2$, discarding $v_{W^1}, v_{W^2}$. Otherwise, return "failure".

---

Figure 4: Find a two-way $\frac{1}{2}$-vertex-separator of $W$ in $G$.

**Lemma 3.7** Let $G(V, E)$ be a graph, $k \geq 0$ an integer, and $W \subseteq V$ of size $3k + 2$. Algorithm $\frac{1}{2}$-vtx-sep($W$, $G$, $k$) finds a two-way $\frac{1}{2}$-separator of $W$ in $G$ of size $\leq 1\frac{1}{2}k$, if it exists, returning failure otherwise. It does so in time $O(\frac{2^{3k}}{\sqrt{k}} f(|V|, |E| + k^2, k))$, given a min-$(a, b)$-vertex-separator algorithm taking time $f(n, m, k)$.

PROOF We prove the correctness of the algorithm. First, assume that the algorithm finds a separator $X$ of $S_1, S_2$ in $G'$. $X$ is also a separator of $S_1, S_2$ in $G$, by the way we constructed $G'$ from $G$. Also, $X$ separates $W^1 \setminus X$ and $W^2 \setminus X$ in $G'$ because $W^1 \cup \{v_{W^1}\}$ and $W^2 \cup \{v_{W^2}\}$ are cliques in $G'$ and $X$ separates $v_{W^1}, v_{W^2}$ (if $X$ does not separate $W^1 \setminus X$ and $W^2 \setminus X$ in $G'$, then there is a path between $v_{W^1}, v_{W^2}$ that does not go through $X$). Finally, $X$ is a $\frac{1}{2}$-vertex-separator of $W$ because $|W^1| = |W^2| = \frac{|W|}{2}$, $W^1 \setminus X \subset S_1$ and $W^2 \setminus X \subseteq S_2$. $S_1, S_2$ are non-empty because $|S_1| \geq |W| - |X| - |W^2| \geq 3k + 2 - 1\frac{1}{2}k - (\frac{1}{2}k + 1) = 1$ (similarly for $S_2$).

Now, assume that there is a two-way $\frac{1}{2}$-vertex-separator $X$ of $W$ in $G$ with $|X| \leq 1\frac{1}{2}k$. Let $S_1, S_2$ be two separated sets of vertices in $G$ such that $S_1 \cup S_2 \cup X = V$ and $|S_i \cap W| \leq \frac{|W|}{2}$, for $i = 1, 2$. Let $W^i_{\text{n\_sep}} = W \cap S_i$, for $i = 1, 2$. Let $W^i_{\text{sep}} \subseteq W \cap X$ such that $W^1_{\text{sep}} \cup W^2_{\text{sep}} = W \cap X$ and $|W^i_{\text{sep}} \cup W^i_{\text{n\_sep}}| = \frac{|W|}{2}$. Let $W^i = W^i_{\text{sep}} \cup W^i_{\text{n\_sep}}$, for $i = 1, 2$. Then, $X$ separates $W^1 \setminus X, W^2 \setminus X$, as $W^i \setminus X = W^i_{\text{n\_sep}}$, for $i = 1, 2$. Thus, running steps 2,3 in our algorithm using this selection of $W^1, W^2$ will find a separator of size $\leq |X| \leq 1\frac{1}{2}k$. By the previous paragraph, this separator is a $\frac{1}{2}$-vertex-separator of $W$ in $G$.

Finally, each choice of $W^1$ takes $O(f(|V|, |E| + k^2, k))$ time to check, for $f(n, m, k)$ the time taken by a min-$(a, b)$-vertex-separator algorithm over a graph with $n$ vertices, $m$ edges and treewidth $k - 1$. There are $\binom{3k+2}{1\frac{1}{2}k+1}$ ways to choose $1\frac{1}{2}k + 1$ elements ($W^1$) from a set of $3k + 2$ elements ($W$). Since $\binom{2n}{n} = \frac{2^{2n}}{\sqrt{\pi n}}(1 + O(\frac{1}{n}))$, we get the time bound of $O(\frac{2^{3k}}{\sqrt{k}} f(|V|, |E| + k^2, k))$. ∎

**Theorem 3.8** Procedure 2way-half-triang($G$, $\emptyset$, $k$) finds a triangulation of $G$ of clique number $\leq 3k + 2$, if the treewidth of $G$ is at most $k - 1$, in time $O(2^{3k} n^{\frac{5}{2}} k^{\frac{1}{2}})$ or $O(2^{3k} n^2 k^{\frac{3}{2}})$ if we use the minimum $(a, b)$-vertex-separator algorithm of [Even, 1979] or [Ford Jr. and Fulkerson, 1962], respectively.

The proof of this theorem is similar to that of Theorem 3.4.

## 4 USING 3-WAY VERTEX SEPARATORS

The last section presented algorithms that recursively divide the set of vertices into two sets. Doing so we give up some of the separators guaranteed by Lemma 2.4. In this section we present a different angle on the tradeoff between the size of the separator, the size of each of the separated sides and the computational complexity of finding the separator. We find approximate *three-way separators*, and use them in a similar way to the one used above.

## 4.1 MULTIWAY VERTEX CUT

A generalization of the minimum $(a, b)$-vertex-cut problem is the minimum multiway-vertex-cut. Given an undirected graph, $G(V, E)$, and a set of nodes, $v_1, ..., v_l \in V$, a minimum multiway cut is a minimum-cardinality set of nodes $S \in V$ such that $v_1, ..., v_l$ are in different connected components in $V \setminus S$. The weighted version requires a minimum-weight set of nodes.

Unlike the minimum $(a, b)$-vertex-cut problem, the problem of finding a minimum multiway-vertex-cut is NP-hard and MAXSNP-hard for $l \geq 3$ [Cunningham, 1991, Garg et al., 1994] (i.e., there is $\epsilon > 0$ such that approximating the problem within a factor of $(1 + \epsilon)$ is NP-hard). For the $(a, b)$-vertex-cut problem the maximum flow is equal to the minimum capacity cut in both directed and undirected graphs. This is not the case for multiway-vertex-cut. Nevertheless, [Garg et al., 1994] showed that by solving a maximum multicommodity flow problem, one can find an $l$-way vertex cut (in an undirected graph) that is of size within a factor $(2 - \frac{2}{l})$ to the optimal (*multicommodity flow* is a generalization of maximum-flow for multiple sources, sinks and commodities sent between them [Leighton and Rao, 1999]).

This algorithm was used subsequently by [Becker and Geiger, 1996] to offer an algorithm for minimum-treewidth triangulation. This algorithm takes time $O(2^{4.66k} n\, poly(n))$, for $poly(n)$ the time required to solve a linear program of size $n$.

## 4.2 FACTOR-$3\frac{2}{3}$ APPROXIMATION ALGORITHM

Figure 5 recalls the main loop of the algorithm of [Becker and Geiger, 1996]. The algorithm differs from that of [Robertson and Seymour, 1995] in using a 3-way separator instead of a 2-way separator. The separator, $X$ of $W$, is required to satisfy $|(S_i \cap W) \cup X| \leq (1 + \alpha)k$, for all three sets $S_i$, $i = 1, 2, 3$, for a given $\alpha \geq 1$. Let us call such a separator a $\alpha$-sum-separator.

Figure 6 presents a new procedure for producing an $\alpha$-sum-separator. It calls a procedure for 3-way vertex separation $3^{|W|}$ times instead of $4^{|W|}$ times as in the algorithm of [Becker and Geiger, 1996].

**Lemma 4.1** *Let $G(V, E)$ be a graph, $k \geq 0$ an integer, and $W \subseteq V$ of size $(1 + \alpha)k + 1$. Algorithm $\alpha$-sum-sep(W, G, k) finds a $\alpha$-sum-separator of $W$ in $G$, if it exists, returning failure otherwise. It does so in time*

---

PROCEDURE 3way-triang($G, W, k$)
$G = (V, E)$ with $|V| = n$, $W \subseteq V$, $k$ integer.

1. If $n \leq (2\alpha + 1)k$, then make a clique of $G$. Return.

2. Let $W' \leftarrow W$. Add to $W'$ vertices from $V$ such that $|W'| = (1 + \alpha)k + 1$.

3. Find $X$, a minimum $\alpha$-sum-separator of $W'$ in $G$, with $S_1, S_2, S_3$ three parts separated by $X$ (at least two are nonempty) and $S_1 \cup S_2 \cup S_3 \cup X = V$. If there is none, then output "*the treewidth exceeds $k - 1$*" and exit.

4. For $i \leftarrow 1$ to 3 do
   (a) $W_i \leftarrow S_i \cap W$.
   (b) call 3way-triang($G[S_i \cup X], W_i \cup X, k$).

5. Add edges between vertices of $W \cup X$, making a clique of $G[W \cup X]$.

Figure 5: A factor-$3\frac{2}{3}$ approximate triangulation algorithm.

---

PROCEDURE $\alpha$-sum-sep($W, G, k$)
$G = (V, E)$ with $|V| = n$, $W \subseteq V$, $k$ integer.

1. Nondeterministically divide $|W|$ into three sets, $W^1, W^2, W^3$, such that $\frac{|W|}{2} \geq |W^1| \geq |W^2| \geq |W^3|$.

2. If $|W^1| > k$, then set $W^2 \leftarrow W^2 \cup W^3$ and return the result of steps 2–3 of algorithm $\frac{2}{3}$-vtx-sep (Figure 2).

3. Let $G' \leftarrow G$. Add edges to $G'$ so that each of $W^1, W^2, W^3$ is a clique. Create new vertices, $v_{W^1}, v_{W^2}, v_{W^3}$ in $G'$ and connect them to all the vertices of $W^1, W^2, W^3$, respectively.

4. Find an $\alpha$-approximation to a minimum $(v_{W^1}, v_{W^2}, v_{W^3})$-vertex-separator, $X$. If $|X| \leq \alpha k$, return $X$ and the three separated sets, $S_1, S_2, S_3$, discarding $v_{W^1}, v_{W^2}, v_{W^3}$. Otherwise, return "*failure*".

Figure 6: Find an $\alpha$-sum-separator in $G$ of size at most $k$.

---

$O(2^{3.6982k} f(|V|, |E| + k^2, k))$, for $\alpha = \frac{4}{3}$, with $f(n, m, k)$ the time taken by an algorithm for $\alpha$ approximation to min-$(a, b, c)$-vertex-separator.

PROOF    We prove the correctness of the algorithm first. Assume that the algorithm finds a $\alpha$-sum-separator $X$ of $S_1, S_2, S_3$ in $G'$. $X$ is also a separator of $S_1, S_2$ in $G$, by the way we constructed $G'$ from $G$. Also, $X$ separates $W^i \setminus X$ and $W^j \setminus X$, $i \neq j \leq 3$, in $G'$ because $W^i \cup \{v_{W^i}\}$ and $W^j \cup \{v_{W^j}\}$ are cliques in $G'$ and $X$ separates $v_{W^i}, v_{W^j}$.

To see that $X$ is an $\alpha$-sum-separator of $W$ we examine two cases. In the first, $|W^1| \leq k$. Thus, $|X| \leq \alpha k$ (otherwise we return "*failure*"). $|(S_i \cap W) \cup X| \leq |W^i| + |X| \leq (1+\alpha)k$, for $i = 1, 2, 3$, because $S_i \cap W \subseteq W^i$. Thus, this is an $\alpha$-sum-decomposition. In the second case, $|W^1| > k$. Thus, $|W^2 \cup W^3| < \alpha k$ because $|W| = (1 + \alpha)k$. Also, $|X| \leq k$ because it was returned by step 3 of algorithm $\frac{2}{3}$-vtx-sep (Figure 2). Thus, $|(S_i \cap W) \cup X| \leq |W^i| +$

$|X| \leq (1+\alpha)k$. Notice that $|S_i \cap W| \geq 1$, for at least two of $i = 1, 2, 3$ (i.e., $X$ does not contain at least two of the $W^i$'s). In the first case this is because $|W^2 \cup W^3| = |W| - |W^1| \geq \alpha k + 1 > |X|$ (we set $|W| = (1+\alpha)k+1$). In the second case this is because $|X| \leq k$, $|W^1| > k$ and $|W^2| = |W| - |W^1| \geq |W| - \frac{|W|}{2} > k$.

For the reverse direction, assume that the treewidth of $G$ is $k-1$. We show that the algorithm finds a suitable separator. Let $S_1, S_2, S_3$ be three sets as guaranteed by Lemma 2.5, separated by $X$ in $G$ such that $S_1 \cup S_2 \cup S_3 \cup X = V$, $|S_3 \cap W| \leq |S_2 \cap W| \leq |S_1 \cap W| \leq \frac{|W|}{2}$ and $|X| \leq k$.

If $|W \cap S_1| \leq k$, then let $W^i_{\text{n\_sep}} = W \cap S_i$, for $i = 1, 2, 3$. Let $W^i_{\text{sep}} \subseteq W \cap X$ such that $W^1_{\text{sep}} \cup W^2_{\text{sep}} \cup W^3_{\text{sep}} = W \cap X$ and $|W^i_{\text{sep}} \cup W^i_{\text{n\_sep}}| \leq \frac{|W|}{2}$. Let $W^i = W^i_{\text{sep}} \cup W^i_{\text{n\_sep}}$, for $i = 1, 2, 3$. Then, $X$ separates $W^1 \setminus X$, $W^2 \setminus X$, $W^3 \setminus X$ because $W^i \setminus X = W^i_{\text{n\_sep}}$, for $i = 1, 2, 3$. Thus, running steps 3,4 in our algorithm using this selection of $W^1, W^2, W^3$ will find a separator of size $\leq \alpha|X| \leq \alpha k$. By the first part of the proof, this separator is a $\alpha$-sum-separator of $W$ in $G$.

If $|W \cap S_1| > k$, then let $W^1_{\text{n\_sep}} = W \cap S_1$ and $W^2_{\text{n\_sep}} = W \cap (S_2 \cup S_3)$. Let $W^i_{\text{sep}} \subseteq W \cap X$, $i = 1, 2$ such that $W^1_{\text{sep}} \cup W^2_{\text{sep}} = W \cap X$ and $|W^i_{\text{sep}} \cup W^i_{\text{n\_sep}}| \leq \frac{|W|}{2}$. Let $W^i = W^i_{\text{sep}} \cup W^i_{\text{n\_sep}}$, for $i = 1, 2$. Then, $X$ separates $W^1 \setminus X, W^2 \setminus X$ because $W^i \setminus X = W^i_{\text{n\_sep}}$, for $i = 1, 2$. Thus, running steps 2,3 of algorithm $\frac{2}{3}$-vtx-sep (Figure 2) using this selection of $W^1, W^2$ will find a separator of size $\leq |X| \leq k$. By the first part of the proof, this separator is a $\alpha$-sum-separator of $W$ in $G$.

Finally, each choice of $W^1$ takes $O(f(|V|, |E| + k^2, k))$ time to check, for $f(n, m, k)$ the time taken by a $\alpha$-approximating *3-way-vertex-separator* algorithm (or a minimum vertex separator algorithm, if it takes more time than the approximate 3-way-vertex-separator) over a graph with $n$ vertices, $m$ edges and treewidth $k - 1$. There are at most $3^{|W|}$ ways to divide $W$ into three sets. Since $|W| \leq (1+\alpha)k+1$, we run a vertex separation algorithm at most $3^{(1+\alpha)k+1} = O(3^{2\frac{1}{3}k}) = O(2^{3.6982k})$ times, for $\alpha = \frac{4}{3}$. Thus, the total time is $O(2^{3.6982k} f(|V|, |E|+k^2, k))$. ∎

**Theorem 4.2 ([Becker and Geiger, 1996])** *If $G(V,E)$ is a graph with $n$ vertices, $k \geq 1$ an integer, $\alpha \geq 1$ a real number, and $W \subset V$ such that $|W| \leq (\alpha+1)k+1$, then 3way-vtx-sep($G,W,k$) triangulates $G$ such that the vertices of $W$ form a clique and such that the size of a largest clique of the triangulated graph $\leq (2\alpha+1)k$ or the algorithm correctly outputs that the cliquewidth of $G$ is larger than $k$.*

Solutions for linear programs of multicommodity flow problems are typically slow. The linear programming subroutine used by the procedure of [Becker and Geiger, 1996] for the subroutine of [Garg et al., 1994] can be replaced by the multicommodity flow algorithm of [Leighton et al., 1995]. This combined algorithm finds a factor-$(1+\epsilon)\frac{4}{3}$ approximation to the optimal 3-way separator in time $O(\epsilon^{-2}nm \, lg^4 n)$, given $\epsilon > 0$. Selecting $\epsilon = \frac{1}{8k}$ guarantees that the separator is in fact a factor-$\frac{4}{3}$ approximation to the optimal (because the separator size is integral).

Using this procedure, the complexity of running the algorithm with $\alpha = \frac{4}{3}$ is $O(2^{3.6982k} n f(n, m+k^2, k)) = O(2^{3.6982k} nk^3 n^2 lg^4 n) = O(2^{3.6982k} n^3 k^3 lg^4 n)$. This is an improvement over the $O(2^{4.66k} n \, poly(n))$ of [Becker and Geiger, 1996], especially because we have reduced the exponential dependency on $k$ by a factor of about $2^k$.

### 4.3 FACTOR-$O(lgk)$ APPROXIMATION

In this section we use a procedure reported in [Leighton and Rao, 1999] for finding factor-$\beta lgk$ approximations ($\beta = 720$) to minimum $\frac{2}{3}$-vertex-separator of $W$ in $G$. This procedure calls a subroutine that solves multi-commodity flow at most $O(|W|)$ times. Using the algorithm of [Leighton et al., 1995] for multicommodity flow it takes time $O(|W|k^3 n^2 lg^4 n)$ for a graph of treewidth $k-1$. The triangulation algorithm uses the main loop of algorithm 3way-triang (Figure 5), replacing steps 2,3 by

2-3. Find $X$, an approximate minimum 3-way $\frac{2}{3}$-vertex-separator of $W$ in $G$, with $S_1, S_2, S_3$ the three parts separated by $X$. If $|X| > \beta k$, then output *"the treewidth exceeds $k - 1$"* and exit.

The combined algorithm for triangulation is used in the same way as before: we call lgk-triang($G, \emptyset, k$).

**Lemma 4.3** *If $G(V,E)$ is a graph with $n$ vertices, $k$ an integer and $W \subseteq V$ such that $|W| = \gamma_1 k \cdot lgk$, for $\gamma_1 = 3\beta$, then lgk-triang($G,W,k$) that uses the procedure of [Leighton and Rao, 1999] for finding 3-way $\frac{2}{3}$-vertex separators either outputs correctly that the treewidth of $G$ is more than $k-1$ or it triangulates $G$ such that the vertices of $W$ form a clique and the clique number of the resulting graph is at most $\gamma lgk$, for $\gamma = 4\beta$.*

PROOF    As in the proofs for previous algorithms, this algorithm either outputs (correctly) that the treewidth exceeds $k$, or a triangulated graph. We show that the clique number of this triangulation is at most $\gamma k lgk$.

First, notice that always $|W| \leq \gamma_1 k lgk$. Initially, $|W| \leq \gamma_1$ by our assumption in the statement of the lemma. As the algorithm is called recursively, $|X| \leq \beta k lgk$ and $|W_i| \leq \frac{2}{3}|W| \leq \frac{2}{3}\gamma_1 k lgk$, by induction. Thus, $|W_i \cup X| \leq \frac{2}{3}\gamma_1 k lgk + \beta k lgk$. Since $\gamma_1 = 3\beta$ we get that $|W_i \cup X| \leq \gamma_1 k lgk$ which concludes the induction step ($W$ in the recursive call to the algorithm is $W_i \cup X$).

Now, let $M$ be a maximal clique. If $M$ contains no vertex of $S_i \setminus W_i$, for $i = 1, 2, 3$, then $M$ contains only vertices of $W \cup X$. Thus, $|M| \leq \gamma_1 k l g k + \beta k l g k = \gamma k l g k$. On the other hand, if $M$ contains a vertex of $S_i \setminus W_i$, then it does not contain any vertex of $S_j$, for $j \neq i$. This is because $X$ vertex-separates $S_1, S_2, S_3$ (any two separated vertices cannot have an edge connecting them). Hence, $M$ is a clique in the triangulation of $G[S_i \cup X]$. By induction we know that $|M| \leq \gamma k l g k$. This proves the lemma. ∎

**Theorem 4.4** *Procedure lgk-triang($G$, ∅, $k$) finds a triangulation of $G$ of clique number $\leq \gamma l g k$), for $\gamma = 4\beta$, if the treewidth of $G$ is at most $k - 1$, in time $O(n^3 l g^4 n\, k^5 l g k)$.*

The proof is similar to those for the previous algorithms.

PROOF    Lemma 4.3 guarantees the correctness of the procedure. The time bound is seen noticing that there are at most $n$ invocations of lgk-triang for a graph of $n$ vertices. ∎

## 5  EXPERIMENTAL RESULTS

We have implemented a constructive variant of algorithms 2way-2/3-triang and 2way-half-triang. Given a graph, $G$, they return a tree decomposition of $G$. The main difference between the description given above and our implementation is that we do not increase the size of $W'$ to be $3k + 2$ in step 2) of Figure 1 (we do not know what $k$ is, a priori). Instead, we gradually increase $W'$'s size during the execution of $\frac{2}{3}$-vtx-sep, until we find a cardinality of $|W'|$ for which a minimum separator has both separated sets non-empty. This is particularly useful when only some of the partitions of the tree decomposition are of size close to the limit.

For our implementation we use an implementation of Chekassky and Goldberg [Chekassky and Goldberg, 1997] for Dinitz's max-flow algorithm. We have experimented with several graphs of various sizes and treewidths that are associated with real-world problems. The results are depicted in Figure 7. They were achieved on a Sun SuperSparc 60. For comparison[1] we ran the implementation of the algorithm of [Shoikhet and Geiger, 1997]. Unfortunately, that algorithm did not return answers for any of these graphs after more than three days. This is not surprising if we compare our theoretical results to those reported in [Becker and Geiger, 1996, Shoikhet and Geiger, 1997]. These algorithms have been tested with graphs of treewidths $\leq 6$, $n \leq 50$, $m \leq 110$ (real-world graphs) and treewidth $\leq 10$, $n \leq 100$ (artificially generated), respectively, an order of magnitude lower than those used here.

---

[1]We could not get the implementation of [Becker and Geiger, 1996].

It is important and interesting to notice that the *min-degree* heuristic [Rose, 1974, Kjaerulff, 1993], which iteratively selects a node that has as few neighbors as possible, makes a clique from the neighbors and removes the node, achieved better tree decompositions than our approximation-guaranteed algorithms on these samples. This heuristic takes between 1 second and 2 minutes on our sample graphs with a sub-optimal implementation, but is not guaranteed to approximate the optimal by a constant factor.

## 6  CONCLUSIONS

We presented four related approximation algorithms for triangulation of minimum treewidth. Two of them (2way-2/3-triang and 3way-triang) are modifications of previous algorithms that improve their running speed by a factor exponential in $k$ and polynomial in $n$. A third algorithm (2way-half-triang) has the best combined $n, k$ time bound known for any constant-factor approximation algorithm. The fourth algorithm (lgk-triang) is the first polynomial-time algorithm for an approximation factor that does not depend on $n$. We showed that our algorithms are efficient enough to solve large problems of practical importance. The results of some of the tree decompositions that we produced are currently being used in reasoning with the HPKB and CYC knowledge bases [Cohen et al., 1999, Lenat, 1995] using algorithms of [Amir and McIlraith, 2000].

#### Acknowledgements

I wish to thank the anonymous UAI reviewers for pointing out the use of Ford-Fulkerson maximum-flow algorithm for the case of finding minimum vertex separators in treewidth-bounded graphs. Work with Sheila McIlraith has raised my interest in the methods that iteratively divided graphs in two/three parts. Kirill Shoikhet allowed me to use his code for optimal triangulation. Daphne Koller and Ben Tasker have given me access to their copy of the CPCS bayes networks. Daphne Koller and Irina Rish have pointed out the heuristics I compared with. I have used an implementation of Dinitz flow algorithm that is due to Andrew Goldberg and Boris Chekasski. This research was supported in part by DARPA grant N66001-00-C-8018 (RKF program).

UAI 2001    AMIR    15

| Graph | Nodes | Edges | Time $4\frac{1}{2}$-apx | Time 4-apx | $4\frac{1}{2}$-apx Width+1 | 4-apx Width+1 | *min-degree* Width+1 |
|---|---|---|---|---|---|---|---|
| CYC1 | 142 | 469 | 1min 2sec | 6min 34sec | 21 | 21 | 14 |
| CPCS1 | 360 | 1036 | 8min 50sec | 1hr 11min | 28 | 26 | 21 |
| CPCS2 | 421 | 1704 | 15min 40sec | 3hr 55min | 33 | 33 | 24 |
| HPKB1 | 446 | 2637 | 2hr 7min | 14hr 13min | 58 | 45 | 37 |
| HPKB2 | 570 | 3840 | 7hr 52min | 5days 23hr | 70 | 60 | 41 |

Figure 7: Graphs, their processing time and the resulting width of the decomposition.